\RequirePackage{snapshot}
\documentclass[10pt,conference]{IEEEtran}

\usepackage[utf8]{inputenc}
\usepackage[T1]{fontenc}
\usepackage{graphicx}
\usepackage[caption=false]{subfig}
\usepackage[usenames,dvipsnames,svgnames,x11names,table]{xcolor}
\usepackage{amsmath,amssymb}
\usepackage{ifthen}
\usepackage{xspace}

\usepackage{algorithm}

\usepackage{algpseudocode}
\usepackage{color}
\usepackage{listings}
\usepackage{pifont}
\usepackage{multirow}
\usepackage{url}
\usepackage{booktabs}
\usepackage[abbreviations]{foreign}
\usepackage[binary-units,per-mode=symbol]{siunitx}
\usepackage{tikz}
\usepackage{flushend}
\usepackage{soul} %
\usepackage{fontawesome}

\usepackage[sort]{natbib}
\setcitestyle{square,numbers}
\usepackage[colorlinks=false,linkcolor=blue,urlcolor=blue,citecolor=blue,bookmarks=false,draft]{hyperref}
\renewcommand*{\UrlFont}{\ttfamily\rm}

\usepackage{setspace}
\usepackage[margin=8pt,skip=0pt,belowskip=0pt,font={small,stretch=0.9},labelfont=bf]{caption}

\newboolean{showcomments}
\setboolean{showcomments}{true}
\ifthenelse{\boolean{showcomments}}
{ \newcommand{\mynote}[3]{
   \fbox{\bfseries\sffamily\scriptsize#1}
   {\small$\blacktriangleright$\textsf{\emph{\color{#3}{#2}}}$\blacktriangleleft$}}}
{ \newcommand{\mynote}[3]{}}

\definecolor{darkgreen}{rgb}{0.3,0.5,0.3}
\definecolor{darkblue}{rgb}{0.3,0.3,0.5}
\definecolor{darkred}{rgb}{0.5,0.3,0.3}

\newcommand\copyrighttext{  \footnotesize \textcopyright 2018 IEEE.
    Personal use of this material is permitted.
    Permission from IEEE must be obtained for all other uses,
    in any current or future media, including reprinting/republishing this
    material for advertising or promotional purposes, creating new collective
    works, for resale or redistribution to servers or
    lists, or reuse of any copyrighted component of this work in other works.
    Presented in the
    \href{http://www.lasid.ufba.br/srds2018/view/index.php}{37th IEEE
      International Symposium on Reliable Distributed Systems (SRDS
      '18)}. The final version of this paper is available under DOI: \href{https://doi.org/10.1109/SRDS.2018.00024}{10.1109/SRDS.2018.00024}}
\newcommand\copyrightnotice{\begin{tikzpicture}[remember picture,overlay]
\node[anchor=south,yshift=10pt,fill=yellow!20] at (current page.south) {\fbox{\parbox{\dimexpr\textwidth-\fboxsep-\fboxrule\relax}{\copyrighttext}}};
\end{tikzpicture}}

\begin{document}

\title{Security, Performance and Energy Trade-offs of\\
Hardware-assisted Memory Protection Mechanisms\\[3mm]\normalsize{\bf (Practical Experience Report)}\\[-8mm]}
\date{}

\author{
\IEEEauthorblockN{Christian Göttel,
        Rafael~Pires,
        Isabelly~Rocha,
        Sébastien~Vaucher,
        Pascal~Felber,
        Marcelo~Pasin,
        Valerio~Schiavoni}
\IEEEauthorblockA{University of Neuch{\^a}tel, Switzerland --- \texttt{first.last@unine.ch}}}

\maketitle
\copyrightnotice

\begin{abstract}
The deployment of large-scale distributed systems, \emph{e.g.}, publish-subscribe platforms, that operate over sensitive data using the infrastructure of public cloud providers, is nowadays heavily hindered by the surging lack of trust toward the cloud operators. 
Although purely software-based solutions exist to protect the confidentiality of data and the processing itself, such as homomorphic encryption schemes, their performance is far from being practical under real-world workloads.

The performance trade-offs of two novel hardware-assisted memory
protection mechanisms, namely AMD SEV and Intel SGX - currently available on the market to tackle this problem, are described in this practical experience.

Specifically, we implement and evaluate a publish/subscribe use-case and evaluate the impact of the memory protection mechanisms and the resulting performance. 
This paper reports on the experience gained while building this system, in particular when having to cope with the technical limitations imposed by SEV and SGX. 

Several trade-offs that provide valuable insights in terms of latency,
throughput, processing time and energy requirements are exhibited by
means of micro- and macro-benchmarks.
\end{abstract}

\section{Introduction}
\label{sec:introduction}

Nowadays, public cloud systems are the \emph{de facto} platform of choice to deploy online services. %
As a matter of fact, all major IT players provide some form of ``infrastructure-as-a-service'' (IaaS) commercial offerings, including Microsoft~\cite{azureconfidential}, Google~\cite{gceskylake} and Amazon~\cite{amazonskylake}.
IaaS infrastructures allow customers to reserve and use (virtual) resources to deploy their own services and data.
These resources are eventually allocated in the form of virtual machines (VMs)~\cite{amazon-ec2}, containers~\cite{google-kub} or bare-metal~\cite{amazon-baremetal} instances over the cloud provider's hardware infrastructure, in order to execute the applications or services of the customers.

Among the many types of communication services, publish/subscribe systems~\cite{eugster2003many} received much attention recently with the objective to support privacy-preserving operations.
Privacy can relate to subscriptions~\cite{barazzutti2012thrifty} (\eg filters that match the subscription of customers to specific pay-per-view TV streaming channels), publisher identities~\cite{raiciu2006enabling} (\eg services providing anonymity to whistleblowers) or the content itself~\cite{nabeel2012efficient}.

These privacy concerns have greatly limited the deployment of such systems over public clouds~\cite{pearson2010privacy}.
Moreover, despite the existence of pure software-based solutions leveraging homomorphic encryption~\cite{naehrig2011can}, their performance is several orders of magnitude behind the requirements of modern workloads.
We evaluated existing homomorphic libraries in order to execute simple operations, such as those typically implemented by publish/subscribe filters, on basic data types.

\begin{figure}[t!]
  \centering  
  \includegraphics[scale=0.7]{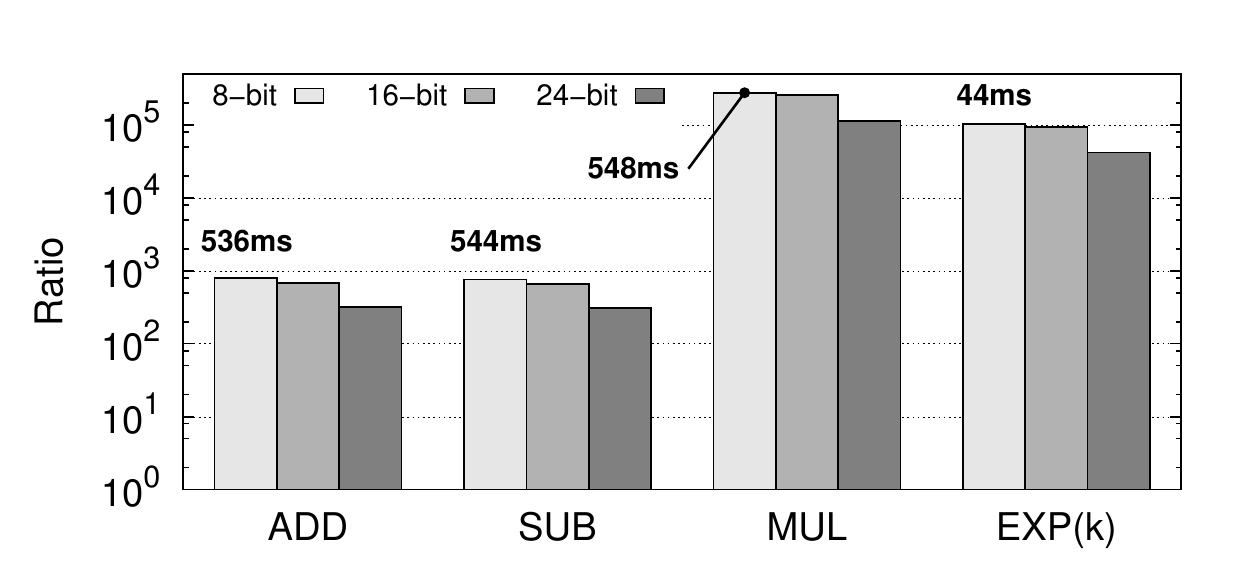}
  \captionsetup{belowskip=-6pt}
  \caption{Performance of simple arithmetic operations using state-of-the-art homomorphic encryption with HElib~\cite{halevi2013design}. Numbers indicate the time to execute a batch of operations in milliseconds.} %
  \label{fig:homomorphic}
\end{figure}

We focused primarily on HElib~\cite{halevi2013design}, which appears to
be the most complete as well as up-to-date, and were able to compare the performance of 8-, 16- and 24-bit addition, subtraction, multiplication and exponentiation operations to a constant value.
\autoref{fig:homomorphic} shows the time ratios for every operation we compared with their unencrypted counterpart, using a \SI{3.1}{\giga\hertz} Intel Core i7 processor with \SI{4}{\mebi\byte} cache (i7-5557U).
For example, the leftmost bar in the figure shows that adding two 8-bit integers with HElib is almost $\num{1000}\times$ slower than adding them unencrypted.
\textsc{Styx}~\cite{Stephen:2016:SSP:2987550.2987574}, an event-based stream processing system that exploits partial homomorphic encryption, confirms our observations.
We can therefore conclude that the performance achievable by these techniques is still unpractical for real-world applications.

The recent introduction of new hardware-assisted memory protection mechanisms inside x86 processors by Intel and AMD paves the way to overcome the limitations of the aforementioned software-only solutions. 
Intel introduced software guard extensions (SGX)~\cite{costan2016intel} with its Skylake generation of processors in August 2015.
These instructions allow applications to create \emph{trusted execution environments} (TEEs) to protect code and data against several types of attacks, including a malicious underlying OS, software bugs or threats from co-hosted applications.
The security boundary of the application becomes the CPU die itself.
The code is executed at near-native execution speeds inside \emph{enclaves} of limited memory capacity.

Along the same line, AMD introduced secure encrypted virtualization (SEV)~\cite{kaplan2016amd,amdsev} with its Zen processor micro-architecture.
Specifically, the EPYC family of server processors introduced the feature on the market in mid-2017~\cite{AMD_press_release:2017,HC29:2017}.
The SEV encrypted state (SEV-ES)~\cite{kaplan2017amd} technology, an extension to SEV, protects the execution and register state of an entire VM from a compromised hypervisor, host OS or co-hosted VMs.
Unmodified applications are protected against attackers with full control over the hosting machine, which in turn can only access encrypted memory pages.

These memory-protection features have been leveraged for several application scenarios where security is paramount.\footnote{We observe that this is particularly true for Intel SGX, since it has been available in the market for much longer. Nevertheless, we expect similar attention to be paid on the AMD platform in the coming months.}
We can cite for example coordination systems~\cite{brenner_securekeeper:_2016}, web search~\cite{mokhtar2017x}, in-memory storage~\cite{enclavedb-a-secure-database-using-sgx}, software-defined networking~\cite{shih2016s}, publish/subscribe systems~\cite{pires2016secure} and streaming platforms~\cite{havet2017securestreams}.
Security researchers have been intensively scrutinizing these features, assessing the claimed security guarantees~\cite{DBLP:conf/vee/HetzeltB17}, discovering new vulnerabilities~\cite{weichbrodt2016asyncshock}, integrating them into container orchestration platforms~\cite{arnautov2016scone}, and evaluating their resilience (or lack thereof) against side-channel attacks~\cite{brasser2017software,lee2017inferring,chen2018sgxpectre}.

However, despite the existing literature, there is no extensive experimental study on the impact of these hardware-assisted memory protection mechanisms for memory-bound applications and systems.
This paper fills this gap by contributing a detailed performance evaluation study, applied to the context of publish/subscribe systems.

The contributions of this paper are as follows.
We explain in detail the differences and similarities, as well as the supported threat models, of the aforementioned hardware architectures.

We detail the engineering efforts in adopting both Intel and AMD hardware solutions (individually).
We evaluate the overhead of SGX and SEV against memory-bound micro-benchmarks.
We execute an extensive evaluation study by means of a complete prototype of an event-based publish/subscribe system.

We finally deploy a realistic scenario and workloads over our publish/subscribe implementation to gather experimental data in real-world settings.

Among the many lessons learned from our experiments, our study suggests that AMD SEV has very little performance impact but, on the other hand, offers weaker security guarantees than Intel SGX.

The remainder of the paper is organized as follows.
Section~\ref{sec:background} provides a technical background on the operating principles for both Intel SGX and AMD SEV.
We then describe the benchmarking architecture (Section~\ref{sec:architecture}) and discuss some implementation details (Section~\ref{sec:implementation}).
Section~\ref{sec:eval} presents a detailed performance comparison between SEV and SGX using micro- and macro-benchmarks.

Section~\ref{sec:rw} surveys related work and finally Section~\ref{sec:conclusion} concludes.

\section{Background}
\label{sec:background}

This section provides background material on the two hardware-assisted memory protection systems that we use in this paper. 
Specifically, \autoref{sec:background:sgx} describes the main concepts and mechanisms of Intel SGX. 
Similarly, \autoref{sec:background:sev} describes how AMD SEV works.

\begin{figure}[!t]
  \centering
  \includegraphics[scale=0.8]{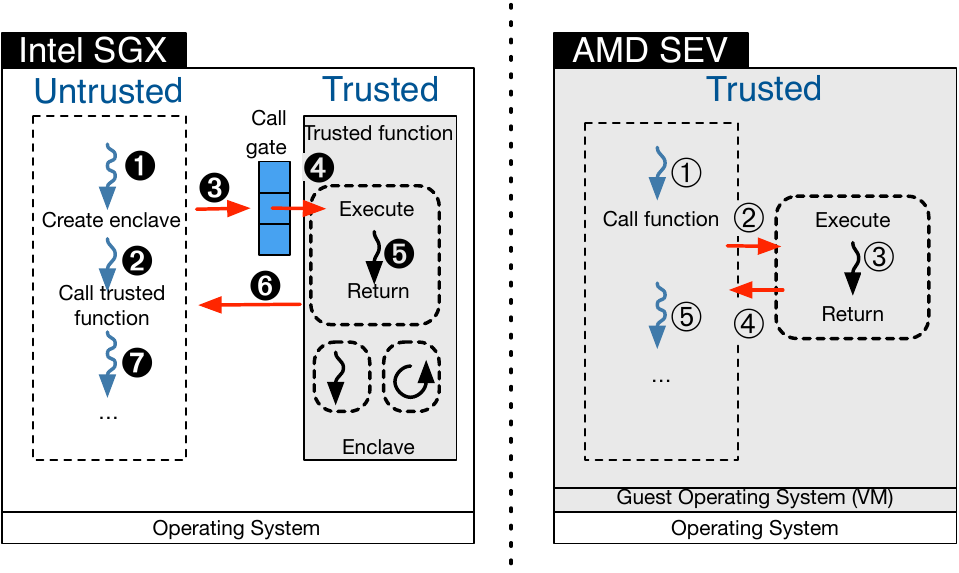}
  \captionsetup{belowskip=-6pt}
  \caption{Intel SGX and AMD SEV operating principles.}
  \label{fig:sgx}
\end{figure}

\subsection{Intel SGX}
\label{sec:background:sgx}

Intel SGX provides a TEE in recent processors that are part of the Skylake and more recent generations.
It is similar in spirit to ARM \textsc{TrustZone}~\cite{arm2009security}.
Applications create secure \emph{enclaves} to protect the integrity and confidentiality of the code being executed and its associated data.

The SGX mechanism, as depicted in \autoref{fig:sgx} (left), allows applications to access confidential data from inside the enclave.
An attacker with physical access to a machine cannot tamper with the application data without being noticed.
The CPU package represents the security boundary.
Moreover, data belonging to an enclave is automatically encrypted and authenticated when stored in main memory.
A memory dump on a victim’s machine will produce encrypted data.
A \emph{remote attestation protocol} (not shown in the figure) is provided to verify that an enclave runs on a genuine Intel processor with SGX enabled.
An application using enclaves must ship a signed, yet unencrypted shared library (a shared object file in Linux) that can be inspected, possibly by malicious attackers.

The \emph{enclave page cache} (EPC) is a \SI{128}{\mebi\byte} area of memory\footnote{Future releases of SGX will relax this limitation~\cite{mckeen2016intel, 2018arXiv180200508K}.} predefined at boot to store enclaved code and data.
At most \SI{93.5}{\mebi\byte} can be used by an application; the remaining area is used to maintain SGX metadata.
Any access to an enclave page outside the EPC triggers a page fault.
The SGX driver interacts with the CPU and decides which pages to evict.
Traffic between the CPU and the system memory is kept confidential by the \emph{memory encryption engine} (MEE)~\cite{gueron2016memory}, also in charge of tamper resistance and replay protection.
If a cache miss hits a protected region, the MEE encrypts or decrypts data before sending to, respectively fetching from, the system memory and performs integrity checks.
Data can also be persisted on stable storage, protected by a seal key.
This allows storing certificates and waives the need of a new remote attestation every time an enclave application restarts.

The execution flow of a program using SGX enclaves is as follows.
First, an enclave is created (see \autoref{fig:sgx}-\ding{202}, left).
As soon as a program needs to execute a trusted function (\ding{203}), it invokes the SGX \texttt{ecall} primitive (\ding{204}).
The program goes through the SGX call gate to bring the execution flow inside the enclave (\ding{205}).
Once the trusted function is executed by one of the enclave's threads (\ding{206}), its result is encrypted and sent back (\ding{207}) before giving back the control to the main processing thread (\ding{208}).

\subsection{AMD SEV}
\label{sec:background:sev}

AMD secure encrypted virtualization (SEV) provides transparent encryption of the memory used by virtual machines.
To exploit this technology, the AMD secure memory encryption (SME) extension must be available and supported by the underlying hardware.
The architecture relies on an embedded hardware AES engine, itself located on the core's memory controller. %
SME creates one single key, used to encrypt the entire memory.
As explained next, this is not the case for SEV, where multiple keys are being generated.
The overhead of the AES engine is minimal. %

SEV delegates the creation of \emph{ephemeral} encryption keys to the AMD secure processor (SP), an ARM \textsc{TrustZone}-enabled system-on-chip (SoC) embedded on-die~\cite{kaplan2016amd}.
These keys are used to encrypt the memory pages belonging to distinct virtual machines, by creating one key per VM.
Similarly, there is one different key per hypervisor.
These keys are never exposed to software executed by the CPU itself.

It is possible to attest encrypted states by using an internal challenge
mechanism, so that a program can receive proof that a page is being
correctly encrypted.%

From the programmer perspective, SEV is completely transparent. 
Hence, the execution flow of a program using it is the same as a regular program, as shown in \autoref{fig:sgx} (right).
Notably, all the code runs inside a trusted environment.
First, a program needs to call a function (\autoref{fig:sgx}-\ding{192}).
The kernel schedules a thread to execute that function (\ding{193}) before actually executing it (\ding{194}).
The execution returns to the main execution thread (\ding{195}) until the next execution is scheduled (\ding{196}). 

\newcommand{\y}{\textcolor{Green4}{\ding{51}}} %
\newcommand{\n}{\textcolor{Red4}{\ding{55}}} %

\begin{table}[!t]
 	\renewcommand{\arraystretch}{0.8}
	\caption{Comparison between Intel SGX and AMD SEV.}
	\label{tab:sgxvssev}
	\setlength{\tabcolsep}{2.3pt}
	\centering
	\rowcolors{1}{gray!10}{gray!0}
	\subfloat[Trusted computing base.]{%
		\resizebox{130pt}{!}{%
			\begin{tabular}{rccc}
				\rowcolor{gray!0}
				\toprule			
				& \textbf{Intel SGX} &
                                \multicolumn{2}{c}{\textbf{AMD}}  \\
                                & & SEV & SEV-ES \\
				\midrule
				Other VMs              & \n   & \n & \n \\
				Hypervisor             & \n   & \y & \n \\
				Host operating system  & \n   & \y & \n \\
				Guest operating system & \n   & \y & \y \\
				Privileged user        & \n   & \y & \y \\
				Untrusted code         & \n   & \y & \y \\
				Trusted code           & \y   & \y & \y \\
				\bottomrule
		\end{tabular}}%
		\label{tab:tcb}
	}
	\subfloat[Features.]{
		\raggedright
		\resizebox{106pt}{!}{%
			\begin{tabular}{rcc}
				\rowcolor{gray!0}
				\toprule
				& \textbf{Intel SGX}   & \textbf{AMD SEV}  \\
				\midrule
				Memory limit        & \SI{93.5}{\mebi\byte} & \emph{n/a} \\
				Integrity           & \y   & \n \\
				Freshness           & \y   & \n \\
				Encryption          & \y   & \y \\
				\bottomrule
			\end{tabular}}
		\label{tab:features}
	}
	\vspace{-10pt}
\end{table}

\subsection{SGX \emph{vs.} SEV}
\label{subsec:sgx-vs-sev}

We briefly highlight the differences between these two technologies along three different criteria, summarized in \autoref{tab:sgxvssev}.

\textbf{Memory limits.} The EPC area used by SGX is limited to \SI{128}{\mebi\byte}, of which \SI{93.5}{\mebi\byte} are usable in practice by applications.
The size of the EPC can be controlled (\ie reduced) by changing settings in the UEFI setup utility from the BIOS of the machine.
This limit does not exist for SEV: applications running inside an encrypted VM can use all its allocated memory.

\textbf{Usability.} To use SGX enclaves, a program needs to be modified---requiring a re-compilation or a relink---\eg using the official Intel SGX SDK~\cite{sgx-sdk}. 
It is the responsibility of developers to decide which sections of the programs will run inside and outside the enclave.
Recently, semi-automatic tools~\cite{lind2017glamdring} have been introduced to facilitate this process. 
As mentioned in the previous section, no changes need to be made to programs when using SEV.

\textbf{Integrity protection.}
Intel SGX has data-integrity protection mechanisms built-in.
Memory pages that are read from EPC memory by an enclave are decrypted by the CPU, and then cached within the processor.
In the opposite direction, data that is being written to the EPC by an enclave is encrypted inside the CPU before leaving its boundaries.
The integrity of the data is safeguarded by associating metadata that is themselves integrity protected.

The metadata is stored in a Merkle tree structure~\cite{1183547}, the root of which is stored in SRAM, inside the processor.
These integrity mechanisms incur an overhead that has been previously evaluated and shown to be acceptable for sequential read/write operations, but up to $10\times$ for random read/write operations~\cite{arnautov2016scone}.

Conversely, to the best of our knowledge, the current version of AMD SEV (or SME) does not provide any integrity protection mechanism~\cite{morbitzer2018severed}.
We expect this limitation to be addressed in future revisions. 

\begin{figure}[t!]
	\centering  
	\includegraphics[scale=0.8]{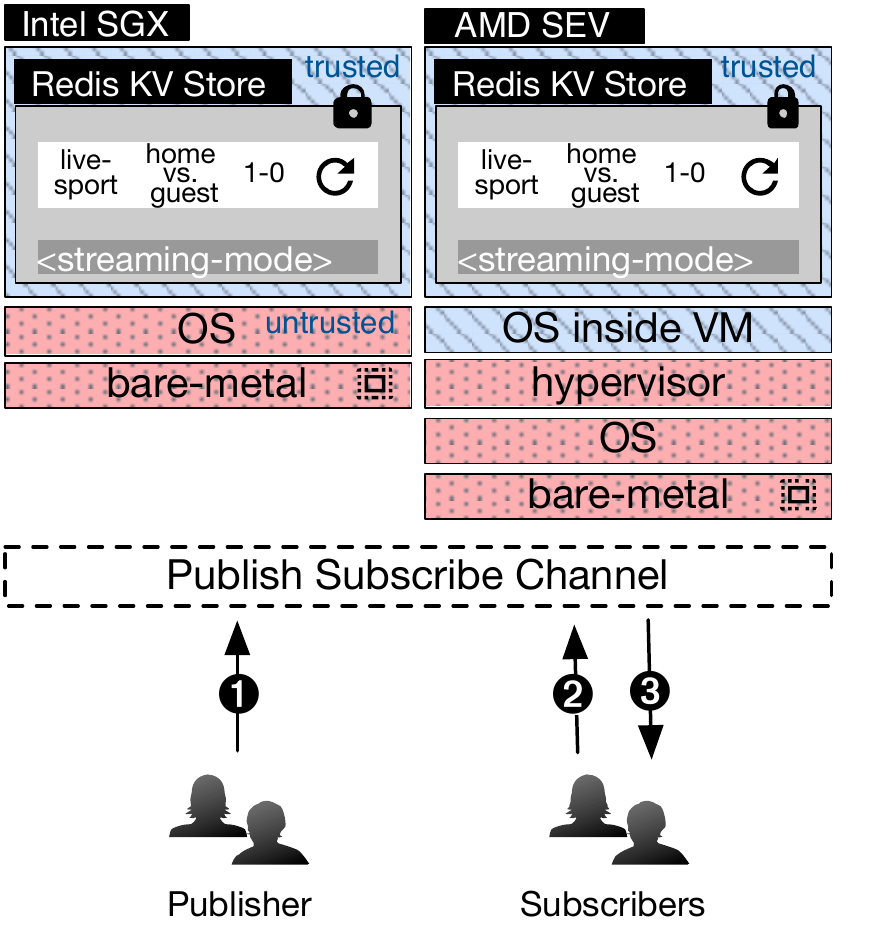}
	\captionsetup{belowskip=-10pt}
	\caption{Architecture of our system and differences when deployed with Intel SGX and AMD SEV-ES. The components with a diagonally hatched pattern on a blue background are trusted, those with a dotted red background are untrusted, respectively. Redis is configured in \emph{streaming-mode}~\cite{redisstreaming}.\label{fig:architecture}}

\end{figure}

\section{Architecture}
\label{sec:architecture}

To execute our evaluation, we designed and implemented a simple yet pragmatic event-based streaming system.
At the core of our system, we rely on a key-value store. 
For every operation occurring on the key-value entries (\ie read, write, update, delete), callback functions associated with these events are automatically triggered.
We assume that the key-value store offers native support to register such callbacks, which can be user- or system-defined, as well as a certain degree of freedom to express the operations and access capabilities that they can achieve.
In the context of a publish/subscribe system, the core role of these functions is to implement matching filters for the subscribers.
Upon execution of such callbacks, all the subscribers on the channel are notified and receive the matching event(s).

\autoref{fig:architecture} depicts the main components of the event-based streaming system.
Specifically, each side of the figure shows which components of the architecture run inside the TEE when using Intel SGX (left) and AMD SEV (right).
The key-value store and its content, the callback functions, as well as the endpoints of the publish/subscribe channels are all potentially sensitive targets; they must hence be protected by SGX or SEV.
However, in our implementation, we only consider the entries of the key-value store to be protected by SGX/SEV. 
Note that solutions to protect the channel endpoints exist~\cite{Aublin:2018:LRS:3190508.3190547} but are not integrated in our prototype.

In our architecture, we do not explicitly include brokers or broker overlays~\cite{eugster2003many}, nor do we include other additional stages in the processing pipeline.
The rationale behind this design choice is to better highlight the side-effects of SGX and SEV on the main processing node in carefully controlled conditions.
Our primary interest lies in the evaluation of memory-bound operations and their energy cost.
We leave as future work the extensions to more sophisticated architectural designs.

The workflow of operations is as follows.
First, a subscriber manifests its interests by subscribing to the channel (\autoref{fig:architecture}-\ding{202}).
Then, publishers start emitting events with a given content, \eg, the results of a sport event (\autoref{fig:architecture}-\ding{203}).
As soon as the content is updated, a callback function is triggered (\autoref{fig:architecture}-\faRepeat).
Finally, the potential subscriber(s) receive the event (\autoref{fig:architecture}-\ding{204}).

\section{Implementation Details}
\label{sec:implementation}

We implemented our architecture on top of well-known open-source systems and libraries.
The key-value store at its core is implemented by Redis~\cite{redis} (v4.0.8), an efficient and lightweight in-memory key-value store.
Redis features a built-in publish-subscribe support~\cite{redis:pubsub}, which we exploit to realize our experimental platform.
The publishers and the subscribers connect to the Redis channels using
Jedis~\cite{jedis} (v2.9.0) Java bindings for Redis.
We further leverage Redis's ability to load external modules~\cite{redis:modules} to implement the callback system described earlier, as well as to be able to serve incoming requests in a multi-thread manner.
While Redis remains a single-threaded system, modules can spawn their own threads. 
We leverage this to improve the throughput of the system and better exploit the multi-core machines in our cluster.
While AMD SEV does not require any change to the system under test, this is not the case for Intel SGX.
For our experiments, we rely on Graphene-SGX~\cite{tsai2017graphene}, a library to run unmodified applications inside enclaves.
In order to use it, one has to write a manifest file where it is defined what resources the enclave is allowed to make use of (shared libraries, files, network endpoints etc.).
This file is pre-processed by an auxiliary tool, which then provides signatures checked by the Graphene loader.
To inject the various workloads, we rely on YCSB~\cite{cooper2010benchmarking}, v0.12.0 commit \texttt{3d6ed690}). %

We intend to release our implementation as open source.\footnote{\url{https://github.com/ChrisG55/streaming}}

\section{Evaluation}
\label{sec:eval}

This section reports the results of our extensive experimental evaluation.
We first describe our evaluation settings and the datasets used in our experiments, before presenting and analyzing in depth the results of the micro- and macro-benchmarks.

\subsection{Evaluation Settings}
\label{subsec:eval-settings}

Our evaluation uses two types of machines.
The Intel platform consists of a Supermicro 5019S-M2 machine equipped with an Intel Xeon E3-1275 v6 processor and \SI{16}{\gibi\byte} of RAM.
The AMD machine is a dual-socket Supermicro 1023US-TR4 machine, with two
AMD EPYC 7281 processors and 8$\times$ \SI{8}{\gibi\byte} of DDR4-2666 RAM.
Both client and server machines are connected on a switched Gigabit network.

The two machines run Ubuntu Linux 16.04.4 LTS.
On the AMD platform, we use a specific version of the Linux kernel based on v4.15-rc1\footnote{\url{https://git.kernel.org/pub/scm/linux/kernel/git/torvalds/linux.git/snapshot/linux-00b10fe1046c4b2232097a7ffaa9238c7e479388.tar.gz}} that includes the required support for SME and SEV.
Due to known side-channel attacks exploiting Intel's hyper-threading~\cite{chen2018sgxpectre}, this feature was disabled on the Intel machine, and so was AMD's simultaneous multithreading (SMT) on the AMD machine.
We use the latest version of Graphene-SGX~\cite{tsai2017graphene},\footnote{\url{https://github.com/oscarlab/graphene/tree/2b487b09}} while we rely on the Intel SGX driver and SDK~\cite{sgx-sdk}, v1.9. 
In order to match the hardware specification of the Intel machine, we deployed para-virtualized VMs on the AMD machine, limited to 4 VCPUs, \SI{16}{\gibi\byte} of VRAM and have access to the host's real-time hardware clock.

The power consumptions are reported by a network-connected LINDY iPower Control 2x6M power distribution unit (PDU).
The PDU can be queried up to every second over an HTTP interface and returns up-to-date measurements for the active power at a resolution of
\SI{1}{\watt} and with a precision of \SI{1.5}{\percent}.

\begin{figure*}[ht!]
	\centering  
	\includegraphics{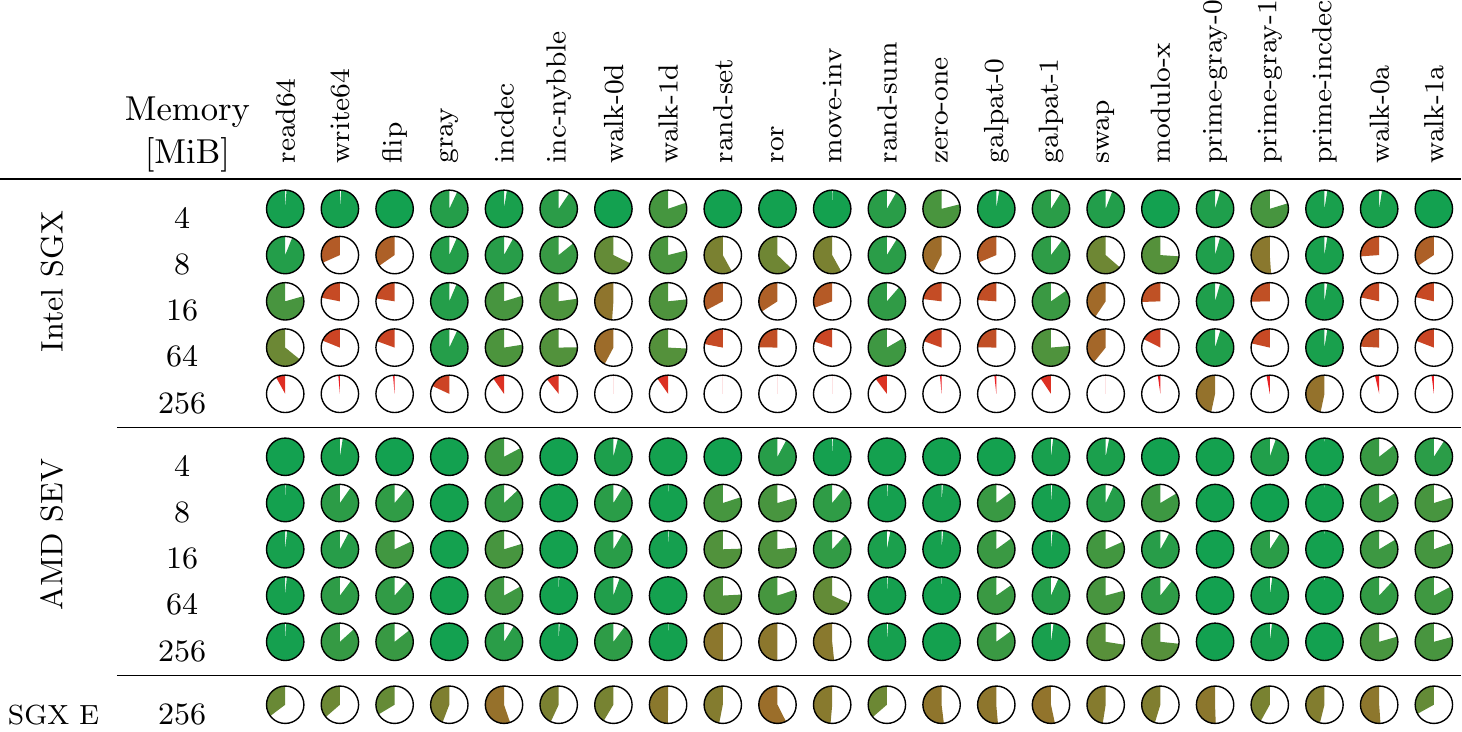}
	\captionsetup{skip=3pt,belowskip=-12pt}
	\caption{Micro-benchmark: relative speed of memory-bound operations using Intel SGX or AMD SEV as protective mechanisms against native performance on each platform. The bottom row shows the relative energy consumption for Intel SGX protective mechanism against native performance. Methods are ordered from sequential (left) to random (right) accesses by increasing memory operation size.\label{fig:micro:mem}}
\end{figure*}

\subsection{Micro-benchmarks}

\textbf{Memory-bound Operations.} We begin with a set of micro-benchmarks to show the performance overhead in terms of memory's access speed imposed by Intel SGX and AMD SEV.
We rely on the virtual memory stressors of \textsc{Stress-NG} as a baseline.
On the Intel architecture, we use \textsc{Stress-SGX}~\cite{stress-sgx}, a fork of \textsc{Stress-NG} for SGX enclaves. 
We ensure that both SGX-protected and unprotected versions of the stressors execute the exact same binary code, to provide results that can be directly compared against one another.

In the case of the AMD machine, the benchmark is first run in a traditional virtual machine, and subsequently the same benchmark is run again with AMD SEV protection enabled. 
We replace the \texttt{mmap} memory allocation functions of the virtual memory stressors with
\texttt{malloc} functions to have a fair comparison between \textsc{Stress-NG} and \textsc{Stress-SGX} (where \texttt{mmap} is not allowed).

\definecolor{piecolor75}{cmyk}{0.69173 0.25436 0.946 0.054}
\definecolor{piecolor100}{cmyk}{0.92233 0.00583 0.928 0.072}
\autoref{fig:micro:mem} summarises the results of this micro-benchmark. 

Values are taken from the average of 10 executions, where each
method is spawning 4 stressors with an execution limit of 30 seconds.
The figure can be read in the following way: the percentage of the surface of each disk that is filled represents the relative execution speed in protected mode, compared to the native speed on the same machine for the same configuration.
For example, a disk that is \SI{75}{\percent} full (\raisebox{-1pt}{\tikz{ \fill[rotate=90, fill=piecolor75] (0,0) -- (0.11,0) arc (0:270:0.11cm) -- cycle; \draw[black, use as bounding box] (0,0) circle (0.11cm); }}) indicates that a stressor ran with protection mechanisms enabled at $0.75\times$ the speed observed in native mode.
A full disk (\raisebox{-1pt}{\tikz{ \draw[black, fill=piecolor100, use as bounding box] (0,0) circle (0.11cm); }}) indicates that the performance of the associated stressor is not affected by the activation of SGX/SEV.

On both platforms, performance is not affected when the program operates on a small amount of memory (\ie, \SI{4}{\mebi\byte}).
The reason is that the protection mechanisms are only used to encrypt data leaving the CPU package.
As \SI{4}{\mebi\byte} is smaller than the amount of cache embedded on the CPU on both platforms (as detailed in \autoref{subsec:eval-settings}), the data never leaves the die and is therefore processed and stored in cleartext.

Both technologies perform better when memory accesses follow a sequential pattern, as observed in the tests \emph{read64, gray, incdec, inc-nybble, walk-d1} and \emph{rand-sum}.
Conversely, Intel SGX is negatively affected by random memory accesses, as seen for tests \emph{swap, modulo-x, prime-gray-1, walk-0a} and \emph{walk-1a}.
AMD SEV is also partially affected under these conditions (tests \emph{swap, modulo-x, walk-0a} and \emph{walk-1a}).
Memory accesses beyond the size of SGX's protected memory (\ie, EPC) are the slowest in our experiment, up to $0.05\times$ less than native memory accesses.
Under these conditions methods such as \emph{modulo-x} were not able to
produce any results. However, supplemental tests, during which
hyper-threading was enabled and all 8 CPUs used, did return results.

\begin{figure*}[!ht]
	\centering
	\includegraphics{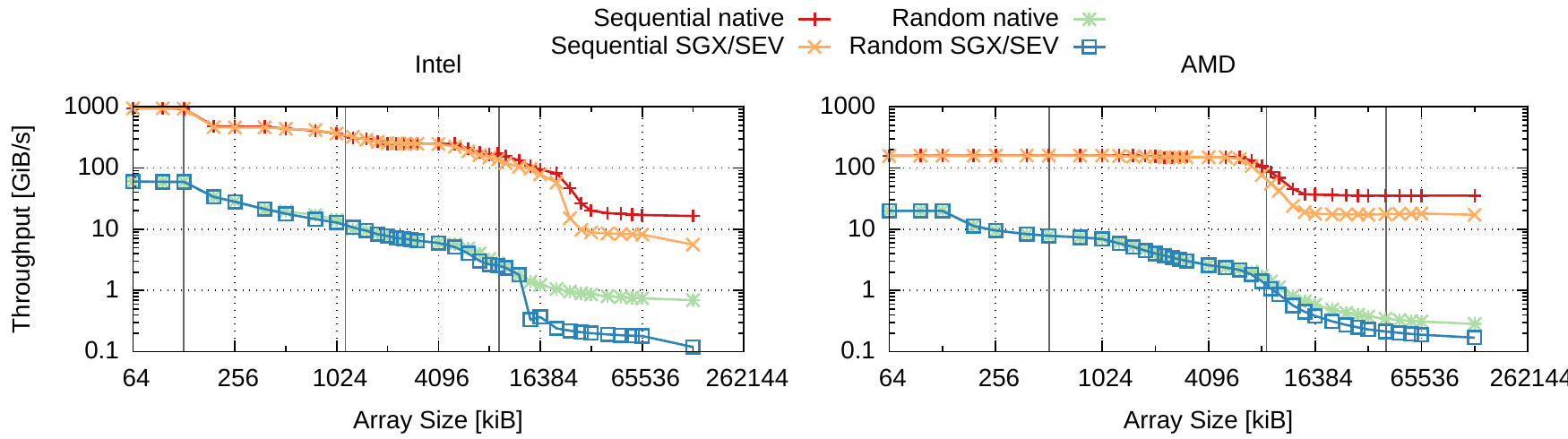}
	\captionsetup{skip=3pt,belowskip=-20pt}
	\caption{Caching effects: measured throughput when accessing a memory region of varying size, sequentially and randomly. Vertical bars highlight the cumulative size of the L1, L2 and L3 caches. Axes are scaled in $\log_2$ and $\log_{10}$. The \emph{qemu} VCPU threads were pinned to a physical CPU on the AMD machine.}
	\label{fig:cache-effect}
\end{figure*}

Finally, SEV appears to be much faster than SGX (an overall \emph{greener} look for the disks), due to its lack of checks to ensure data integrity protection (as explained in \autoref{subsec:sgx-vs-sev}).
Similarly, larger memory accesses also do not suffer from drastic performance penalties like in the case of Intel SGX.

\begin{figure*}
	\centering
        \includegraphics{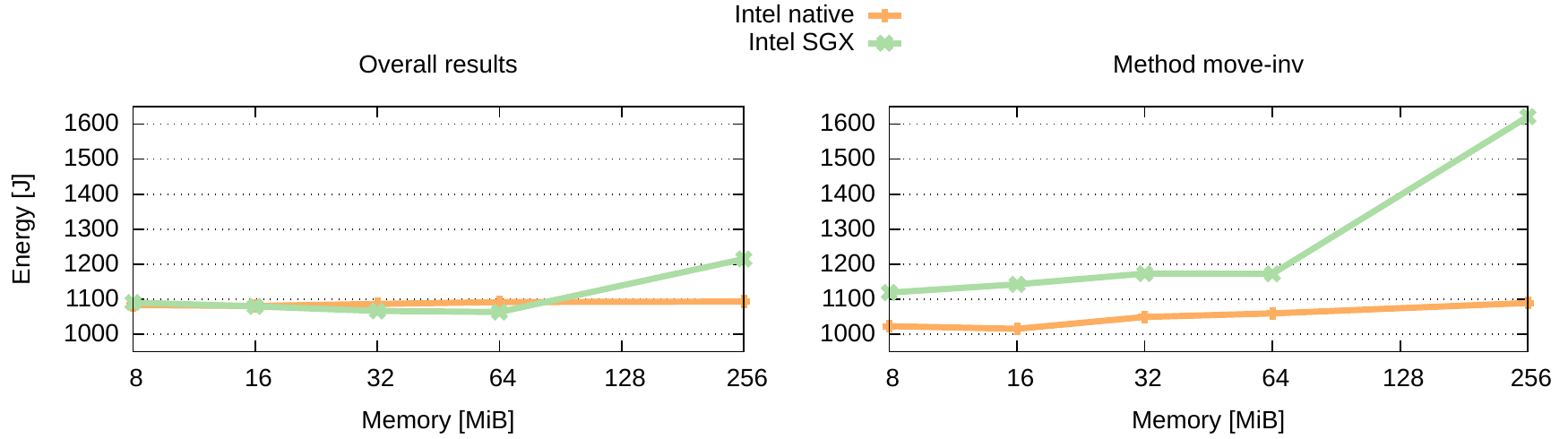}
	\captionsetup{belowskip=-16pt}
	\caption{Energy measurements for micro-benchmark on Intel Xeon
          E3-1275 v6.}
	\label{fig:sgx:native:energy}
\end{figure*}

\textbf{Energy cost of memory-bound operations.} To evaluate the energy cost of memory-bound operations we recorded the
power consumption while running the micro-benchmark of \autoref{fig:micro:mem}.

\definecolor{piecolorinverted}{rgb}{0.67 0.308 0}
The results are shown in the bottom row of the table, row \textit{SGX E}.
The pie-chart is read as follows: a disk that is
\SI{67}{\percent} full (\raisebox{-1pt}{\tikz{ \fill[rotate=90, fill=piecolorinverted] (0,0) -- (0.11,0) arc (0:240:0.11cm) -- cycle; \draw[black, use as bounding box] (0,0) circle (0.11cm); }}) indicates that the \textsc{stress-sgx} method consumed $1.67\times$ more energy during execution with SGX enabled compared to native performance.

As expected, the energy consumption using SGX increases when the memory size considered is bigger than the EPC memory, and a similar behaviour is observed for each of the stressor method.
However, the case of the \emph{move-inv} stressor is different.
In this case (\autoref{fig:sgx:native:energy}\;(right)), SGX mode consumes more energy than native, independently from the memory size.

The \emph{move-inv} stressors sequentially fill memory with random data, in blocks of 64 bits.
Then they check that all values were set correctly.
Finally, each 64 bit block is sequentially inverted, before executing again a memory check.

Conversely, in the case of AMD SEV we did not observe higher energy consumptions compared to native energy consumption, hence these results do not appear in \autoref{fig:micro:mem}.
Specifically, 108 out of 110 memory stressors confirm that the energy consumption lies within the \SI{3.7}{\percent} margin of error, \emph{i.e.}, the precision of the measurement.
Only two measurements (\emph{read64} with memory size \SI{16}{\mebi\byte}, and \emph{modulo-x} with memory size \SI{256}{\mebi\byte}) lie slightly outside the range of error and do not confirm the observation.

\textbf{Caching Effects.} With both AMD SEV and Intel SGX protection mechanisms, data is only encrypted when it leaves the processor package.
In order to show the impact of caching on performance, we measure the throughput for varying sizes of memory accesses. %
We use the \texttt{pmbw} tool~\cite{pmbw} (v0.6.2 commit \texttt{fc712685}) to conduct this experiment, ported with Graphene-SGX~\cite{tsai2017graphene} to run inside an SGX enclave.
To provide a comparable baseline, we also use Graphene~\cite{Tsai:2014:CSI:2592798.2592812} to run the native case on the Intel platform.
On the AMD platform, we run the same micro-benchmark in a virtual machine, with and without SEV enabled.
The \texttt{qemu} VCPU threads were pinned to physical CPUs on the same node in order to augment caching effects exercised during the micro-benchmark.

\autoref{fig:cache-effect} shows the observed throughput averaged over 10 runs when reading through a fixed amount of memory.
The results are presented in the form of a \emph{log-log} plot to clearly highlight the behaviour at each step of the memory hierarchy (L1/L2/L3 caches and main memory).
We see that, within the cache, the performance of both AMD SEV and Intel SGX is strictly equivalent to native performance, in particular within the L2 cache. 
AMD SEV shows some overhead with L3 cache for sequential and random accesses.
When the amount of memory to read surpasses the cache on Intel SGX, the throughput is greatly affected.
As previously reported~\cite{arnautov2016scone}, random accesses to the EPC incur a greater overhead than sequential reads.
In the case of AMD SEV, only a very small overhead can be observed.

We made a couple of surprising observations when running these experiments.

First, on the Intel platform, the significant drop in performance beyond the cache limits happens when the tested memory size is already markedly larger than the total cache size.
This undocumented behavior, although does not affect the final outcome of our study, might be caused by Intel's \emph{smart cache} technology~\cite{intelsmartcache}. %

Second, on the AMD platform, L3 cache performance is decreasing at a much faster rate than on the Intel platform.
This behavior is observed when running both in native and shielded mode.
We assume that the performance decrease is due to the virtualization process of the VM. %
In both cases a more thorough investigation has to be conducted to explain our observations.

\subsection{Macro-benchmarks}
\label{subsec:eval-workloads}

\begin{figure}[t!]
	\centering  
	\includegraphics[scale=0.64]{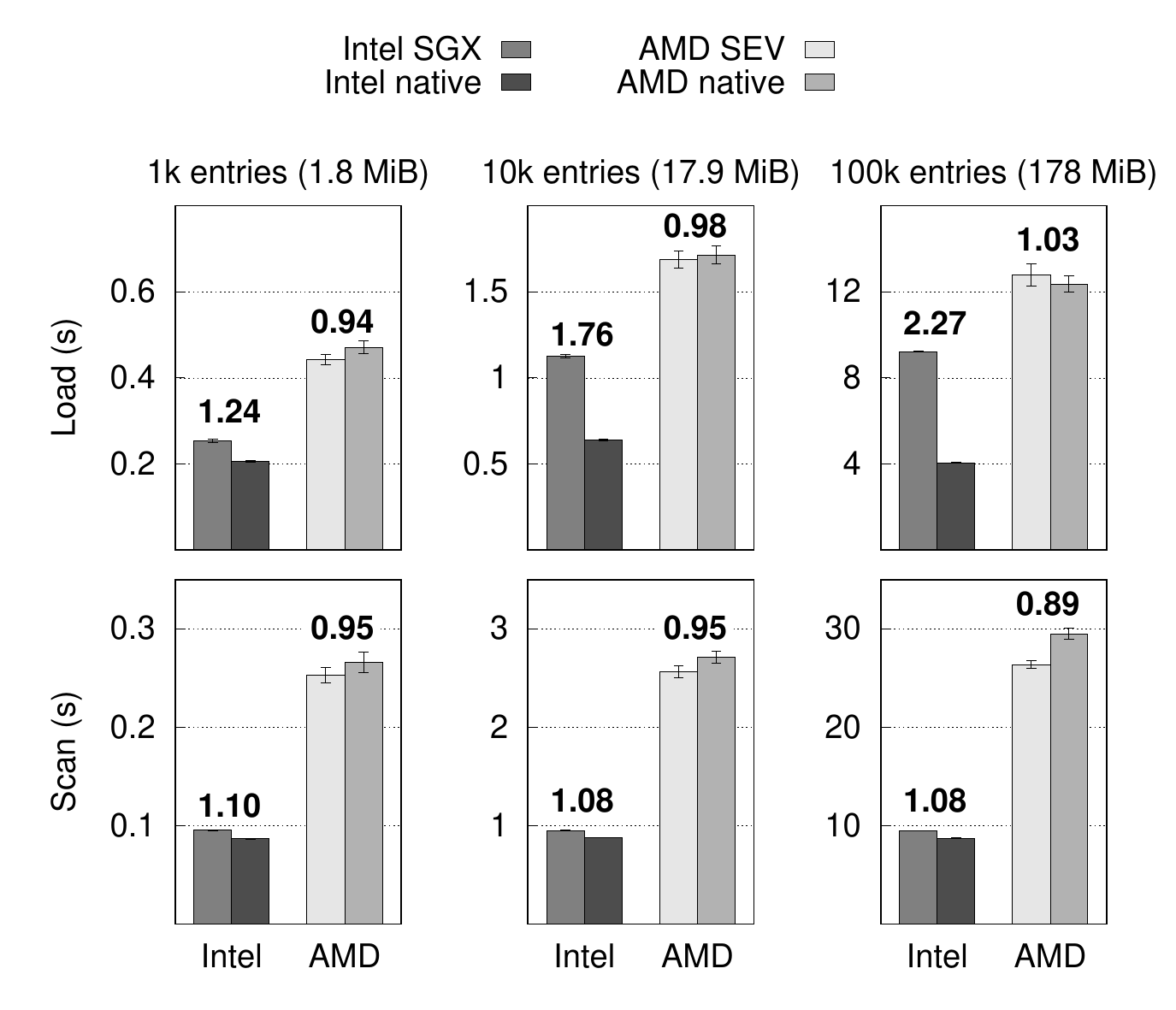}
	\captionsetup{skip=-3pt,belowskip=-6pt}
	\caption{Macro-benchmark: performance of Redis' \emph{load} and \emph{scan} operations for different number of entries.}
	\label{fig:scan}
\end{figure}

\textbf{Workload Description.} We use a simple update-only workload for our first macro-benchmark~(\autoref{fig:scan}).
In order to simulate incremental changes for certain entries in the dataset, we replaced their associated write commands (Redis' \texttt{HMSET}~\cite{redis:hmset}) with update commands (\texttt{HINCRBY}~\cite{redis:hincrby}).
Only a subset of entries selected by a Zipf distribution were replaced.

The second workload is based on the update-heavy YCSB's \emph{workload~A}~\cite{ycsb:workloada}, executed in two phases.
In a first phase, the YCSB loader writes a fixed number of datasets into the Redis database.
Secondly, YCSB runs the benchmark with a number of operations equal to the number of datasets loaded in the former phase.
Operations are issued with a 50/50 read/update split.
A read operation will always read all fields of a Redis Hash, the Redis data type used by YCSB to store datasets.
Datasets to be updated are chosen following a Zipf request distribution.
Finally, we modified and extended the default behaviour of \emph{workload~A} as follows:

\emph{(i)}~the fields of all hashes are limited to a set length;
\emph{(ii)}~small unique identifiers (36 Bytes) are inserted into hash fields to track the events' in-flight time from the publisher to the subscriber(s), \ie, for recording purposes;
\emph{(iii)}~we implemented a specialized word-counting workload which could be used for similarity analysis of documents; and

\emph{(iv)}~operations are issued at a fixed throughput, rather than on a best-effort basis, to evaluate experimentally the saturation point of the system, that is when the latencies increase beyond usable thresholds.

\textbf{Redis LOAD and SCAN.} To evaluate how memory usage impacts the system, we measured the time it takes to accomplish two operations in the Redis server: loading data to the in-memory database, and performing a full iteration for key retrieval by using the Redis \texttt{SCAN} command~\cite{redis}.
The two operations were subsequently executed and independently measured.
We used YCSB \emph{workload~A} dataset (as described in \autoref{subsec:eval-workloads}) with varying number of records, corresponding to 3 scenarios:
\emph{(i)}~below L3 cache size both in Intel and AMD;
\emph{(ii)}~above Intel L3 cache threshold and below AMD's; and
\emph{(iii)}~above SGX EPC limit in Intel and L3 cache in AMD.
For Intel executions, we used Graphene~\cite{Tsai:2014:CSI:2592798.2592812,tsai2017graphene} both in SGX and native scenarios.
\autoref{fig:scan} shows the results for \num{20} executions of each experiment. 
The ratio between each pair of bars is indicated above them.
Error bars correspond to the \SI{95}{\percent} confidence interval assuming a Gaussian distribution.

Looking at Intel SGX results, we clearly notice the evolution of performance drops in the \emph{load} experiment.
The overhead caused by SGX ranges from \SI{24}{\percent} before reaching the L3 cache limit, to \SI{76}{\percent} past the limit.
Likewise, paging brings a considerable cost of \SI{127}{\percent}.
In the \emph{scan} experiment, on the other hand, we see a steady overhead of about \SI{10}{\percent} across different memory sizes, growing strictly linearly in terms of absolute delays.
This might be explained by memory prefetching, since the execution is sequential.
Similar results were obtained in \autoref{fig:cache-effect} and \cite{arnautov2016scone}.

\label{subsec:tputlatency}
\begin{figure}[t!]
  \centering
  \includegraphics[scale=0.7]{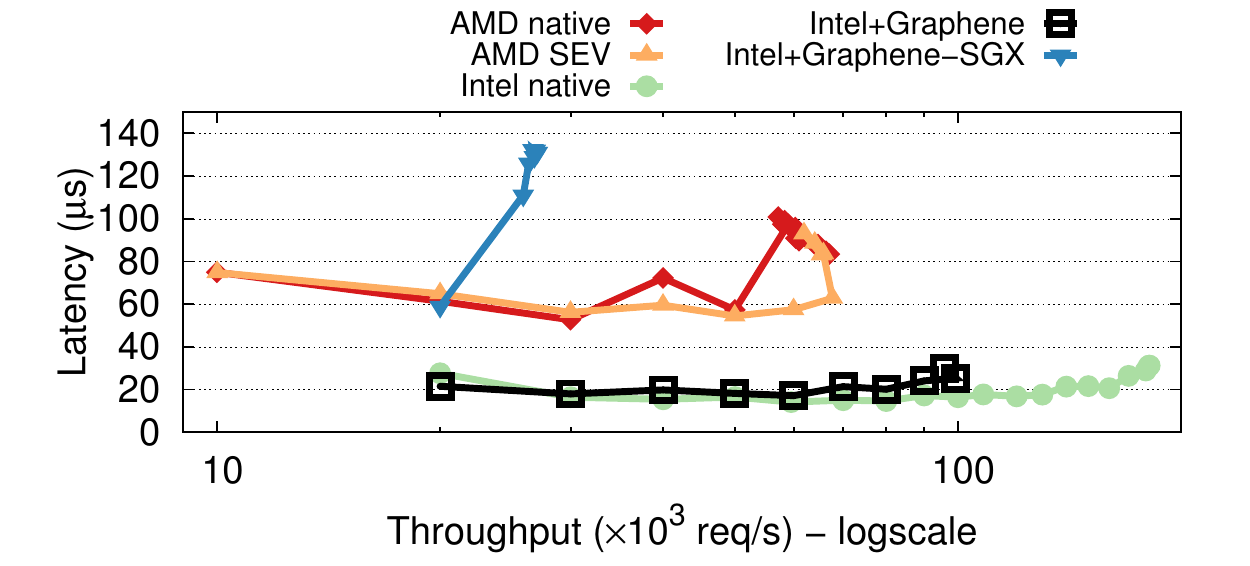}
  \captionsetup{belowskip=-6pt}
  \caption{Macro-benchmark: throughput vs. latency for standalone Redis server with different deployment scenarios (Native Intel, Graphene, Graphene-SGX, Native AMD and AMD SEV.}
  \label{fig:tputlat:kv}
\end{figure}

With regards to AMD experiments, the same unsettling results appeared: looking only at averages, we notice a slight improvement in performance when SEV is turned on, except for 100k entries in the \emph{load} observation. 
Considering the confidence interval, however, they are essentially equal excluding the \emph{scan} experiment for 100k entries.
We plan to further investigate this behaviour.

\begin{figure*}
	\centering
	\subfloat{%
		\includegraphics[width=\columnwidth]{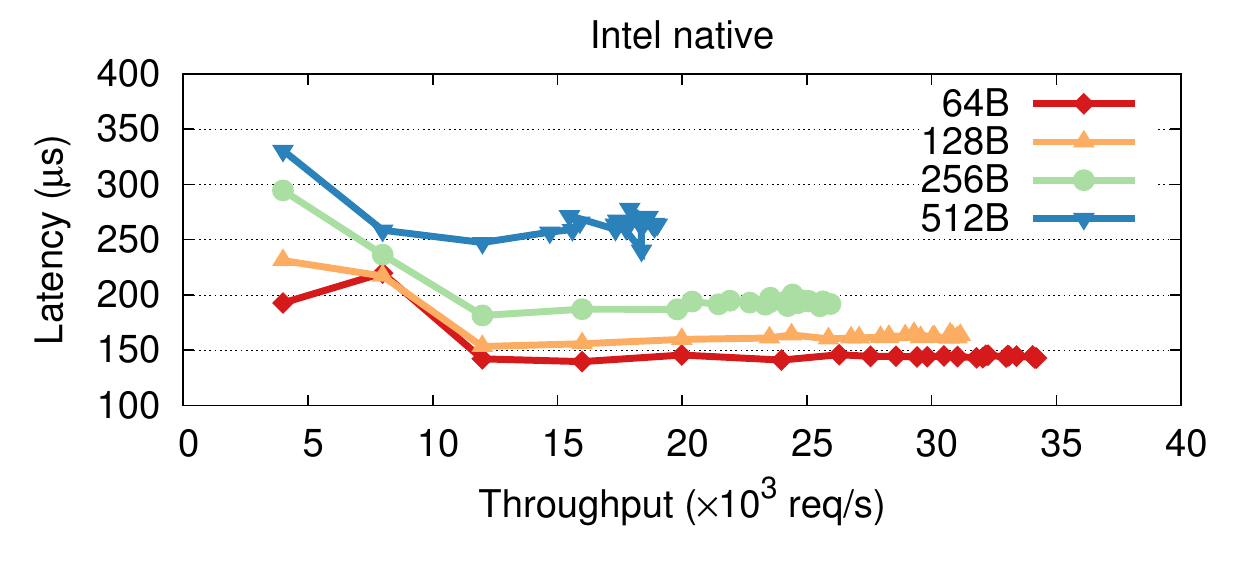}
		\label{fig:tputlat:pubsub:intel}
	}%
	\subfloat{%
		\includegraphics[width=\columnwidth]{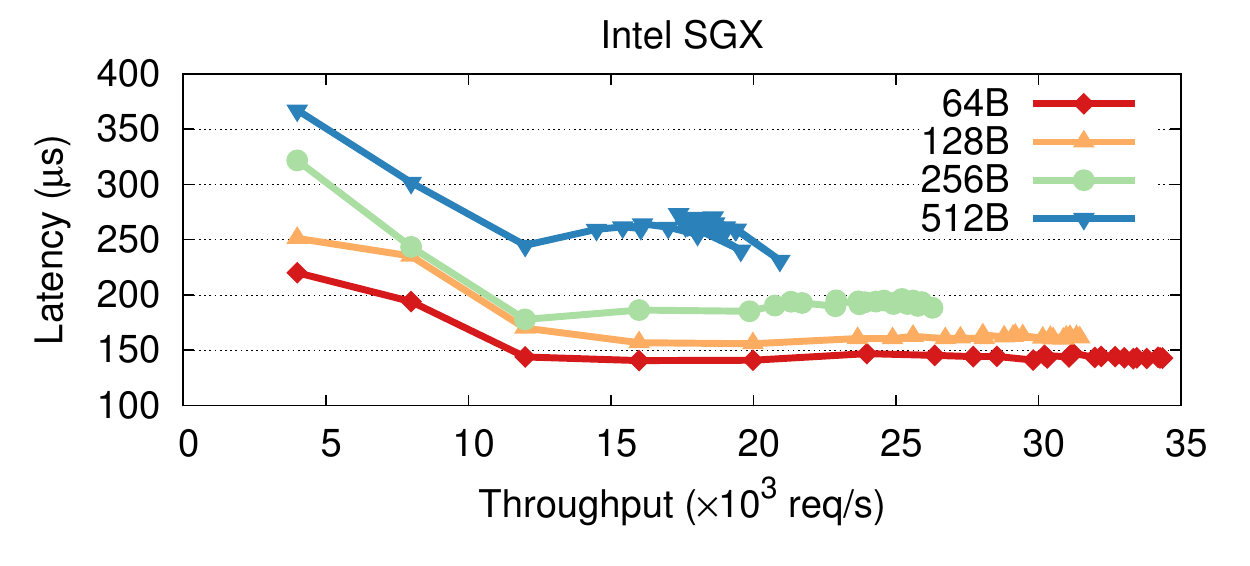}
		\label{fig:tputlat:pubsub:sgx}
	}
	\\[-12pt]
	\subfloat{%
		\includegraphics[width=\columnwidth]{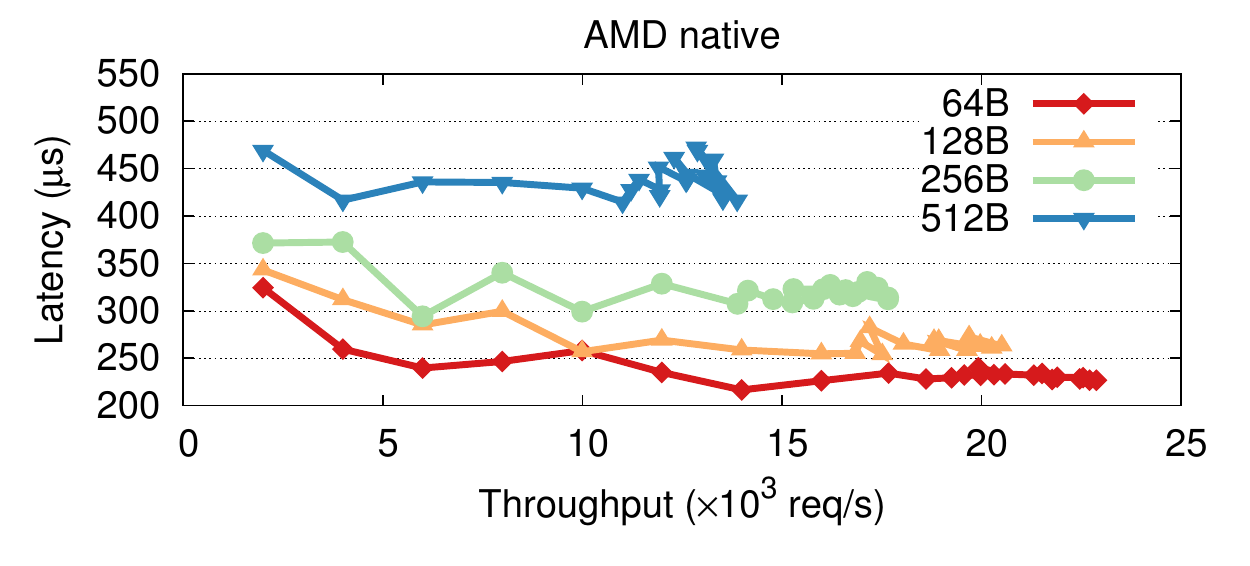}
		\label{fig:tputlat:pubsub:amd}
	}%
	\subfloat{%
		\includegraphics[width=\columnwidth]{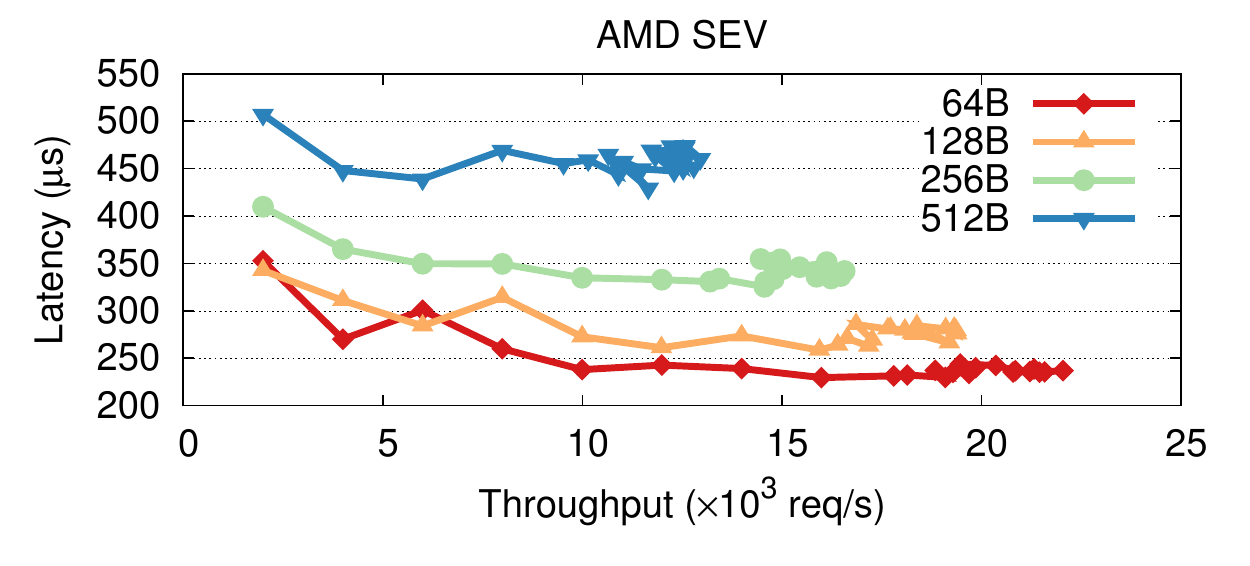}
		\label{fig:tputlat:pubsub:sev}
	}
	\captionsetup{skip=-3pt,belowskip=-12pt}
	\caption{Macro-benchmark: throughput vs. latency.}
    \label{fig:tput}
\end{figure*}

\textbf{Throughput/Latency.} Our second macro-benchmark evaluates how SGX and SEV affect the observed latency of the requests served by Redis.
We incrementally augment the amount of requests per second issued by the client, until the measured latencies spike.
We compare these results against three different baselines, respectively those with the Redis server running on the unshielded AMD machine, the unshielded Intel machine, and when using Graphene but without SGX support.
\autoref{fig:tputlat:kv} shows our results.
On the x-axis (log scale), we report the number of operations per second (scaled by a factor of \num{1000}), and on the y-axis the measured response latency as reported by YCSB.
As expected, the unshielded execution on the Intel machine outperforms the alternatives.
We also observe that the overhead of Graphene itself is modest.
On the other hand, we observe deteriorated results for all the shielded executions.

\textbf{Publish/subscribe.} Our final set of macro-benchmarks deploy the full pub/sub architecture.
The experiments measure the message latency from the moment a publisher emits a new event until the moment all the subscribers receive its content.

Then, we configure the publisher to inject new events at fixed rates.
We evaluate the performance of the system against 4 different configurations (\autoref{fig:tput}):
\emph{(i)}~Intel without SGX protection;
\emph{(ii)}~with SGX by leveraging Graphene;
\emph{(iii)}~AMD without memory protection; and
\emph{(iv)}~AMD with SEV.
For the different configurations we issue requests of 4 different sizes, from \SI{64}{\byte} up to \SI{512}{\byte}, as well as fixed throughputs (on the x-axis of each sub-plot).

We observe that for smaller message sizes, the measured latencies are consistently lower for higher throughputs (requests/second).
With bigger messages, our implementation is less efficient.
This is due to the cost of serializing messages. %
Nevertheless, when doing a pairwise comparison between the Intel and AMD configurations, it is clear how these protection mechanisms are negatively affecting the observed latencies.
This is particularly evident for the Intel configurations.
The bandwidth-wise computations (not shown in \autoref{fig:tput}, calculating how much data is being transfered for each curve) confirm these observations. %

\begin{figure*}
	\centering
	\subfloat{%
		\includegraphics[width=\columnwidth]{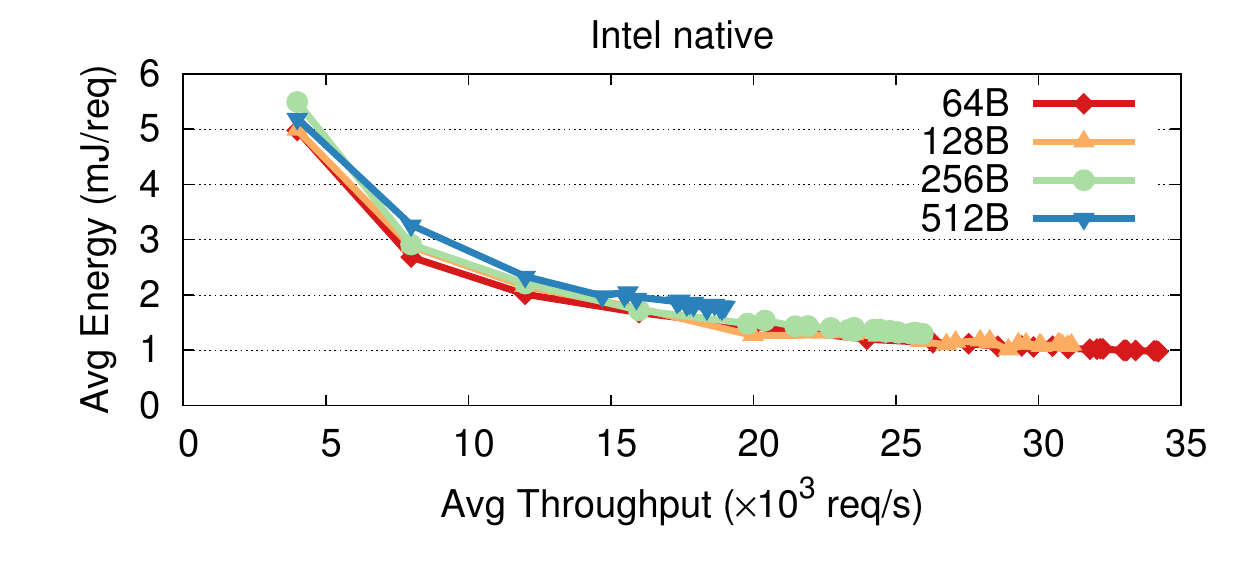}
		\label{fig:pubsub:energy:intel}
	}%
	\subfloat{%
		\includegraphics[width=\columnwidth]{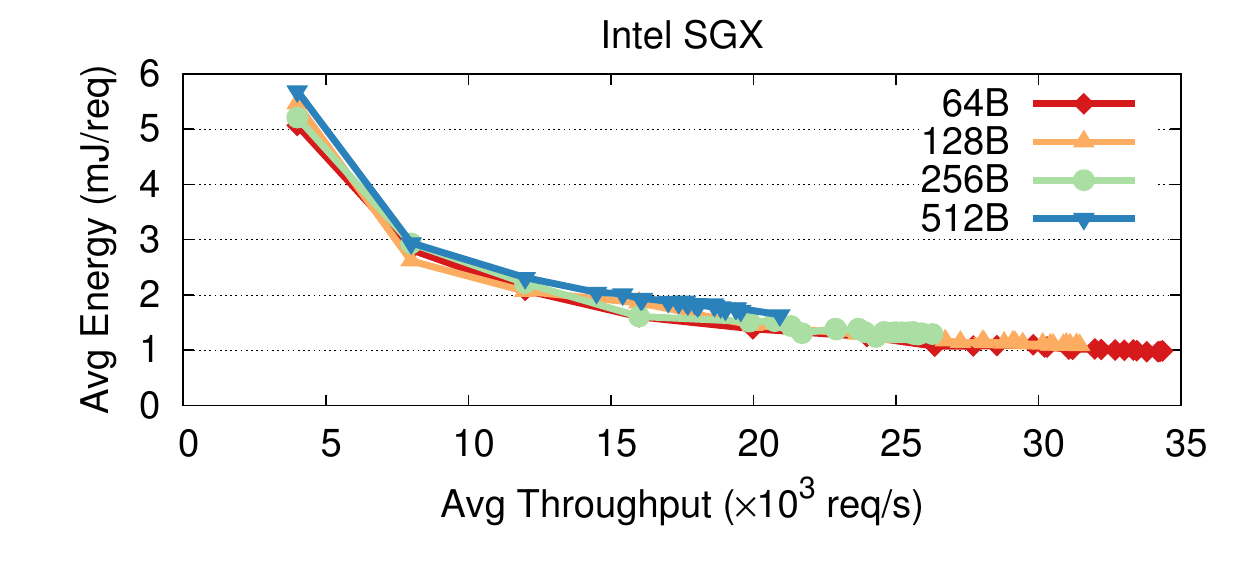}
		\label{fig:pubsub:energy:sgx}
	}
	\\[-12pt]
	\subfloat{%
		\includegraphics[width=\columnwidth]{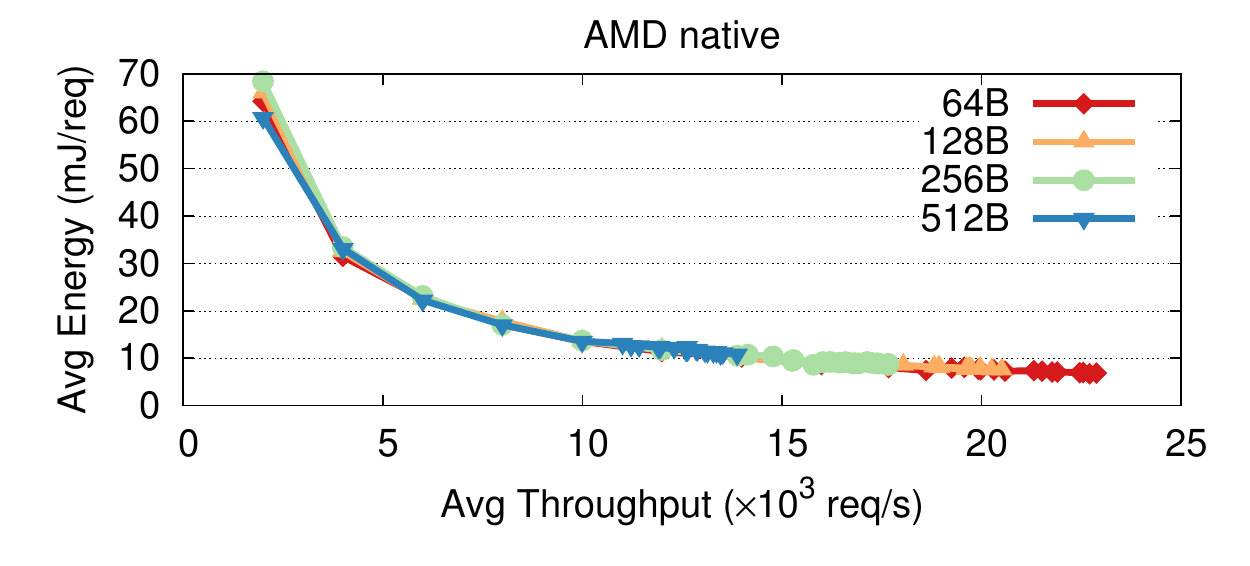}
		\label{fig:pubsub:energy:amd}
	}%
	\subfloat{%
		\includegraphics[width=\columnwidth]{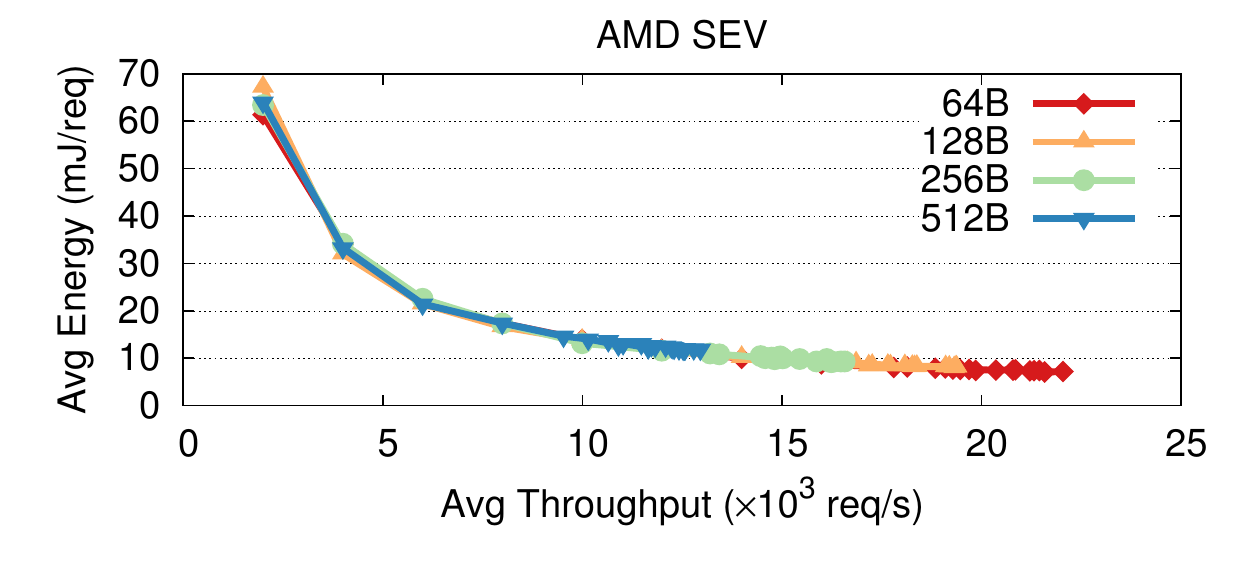}
		\label{fig:pubsub:energy:sev}
	}
	\captionsetup{skip=-3pt,belowskip=-12pt}
	\caption{Macro-benchmark: energy cost of publish/subscribe.}
    \label{fig:pubsub:energy}
\end{figure*}

\textbf{Energy cost of publish/subscribe.} We also recorded the power consumption of the publish/subscribe system shown in Figure~\ref{fig:pubsub:energy}.
Our analysis indicates that the energy consumption increases at a
  linear rate relative to the target throughput once the system begins
  occupying a significant amount of the machine's resources.
This is reflected by the decreasing energy cost per request before
  reaching its minimal cost.

Under these settings, the memory requirements do not exceed the available EPC. %

Hence, both protection mechanisms have a similar energy consumption to their native setup.

It should be noticed that the reported energy consumption includes all components of the machines which comprises auxiliary devices such as the network card.
The results of the macro-benchmark therefore have no direct implication on the energy consumption of the protection mechanisms.
In the future we would like to be able to analyze the energy consumption of processes at a much more fine grained level such as the processor's core.
This will allow us to observe in more detail what impact protection mechanisms exert on processes.

\section{Related Work}\label{sec:rw}

Present solutions for databases and publish/subscribe systems often expose
a significant lack of performance when leveraging software-based
privacy-preserving mechanisms.

Subsequently exemplified solutions could benefit from
hardware-assisted TEEs such as Intel SGX or AMD SEV.

EnclaveDB~\cite{enclavedb-a-secure-database-using-sgx} proposes to run a database engine inside an SGX enclave.
The database engine is split into many components, with an arguably small enclaved component that only stores data considered as sensitive in the enclave.
Yet, experiments have shown that EnclaveDB imposes a \SI{40}{\percent} overhead when compared to a similar native implementation.
EnclaveDB performance figures include memory throughput, based on simulated memory encryption, where enclaves can have up to \SI{192}{\gibi\byte} of memory.
In our work, we include memory throughput figures measured on actual Intel SGX hardware and compare it to AMD SEV.

Merkle hash trees are used to guarantees data integrity to clients for the key-value store VeritasDB~\cite{veritasdb}.
A proxy is implemented to verify the database integrity, intermediating all exchanges between client and server.
The proxy executes inside a trusted enclave, which removes the need for trust on the server.
Although it would be interesting to evaluate how VeritasDB's proxy would perform when running on AMD SEV, our implementation is small enough so the trusted component fits within current SGX EPC limits, eliminating the need for a proxy.

The choice of using Redis as back-end storage in our centralized publish/subscribe framework is shared by other systems.
For instance, Redis was used as well to implement a low-footprint pub/sub framework~\cite{Abidi:2013:PIS:2480362.2480510} for managing a resource-constrained grid middleware.
Also, in DynFilter~\cite{Gascon-Samson:2015:DLB:2984075.2984077}, Redis was used to implement a game-oriented message processing middleware that adaptively filters state update messages.
Similarly, we use Redis and its built-in publish/subscribe capability because of its lightweight implementation, a primary requirement when dealing with the limited amount of EPC available to SGX systems.
Unlike this previous work, we concentrate on the evaluation of a pub/sub engine under trusted execution environments.

PP-CBPS~\cite{nabeel2012efficient} is a content-based publish/subscribe engine based on Paillier's homomorphic encryption.
It shows that it is possible to match a few dozen encrypted publications per second when having a few thousand subscriptions.
On the other hand, high-performance publish/subscribe engines such as StreamHub~\cite{barazzutti2013streamhub} can match tens of thousands plaintext publications per second in similar conditions.
As shown on simple operations in the introduction of this paper, homomorphic encryption still imposes a large performance penalty.

TrustShadow~\cite{Guan:2017:TSE:3081333.3081349} isolates standard applications from the operating system using ARM TrustZone~\cite{arm2009security}, which is to some extent similar to SGX and SEV.
Along the same lines as our SGX approach, TrustShadow executes standard applications inside a trusted environment and coordinates the communication between the application and the operating system.
TrustShadow exercises the processor with a different set of benchmarks, yet it complements our efforts and brings light into the performance effects of using an ARM architecture.

\section{Conclusion}
\label{sec:conclusion}

Privacy-preserving publish/subscribe systems would dramatically benefit from the new wave of trusted hardware techniques that are now available in most recent processors sold by Intel and AMD.
As a matter of fact, their design could be greatly simplified, for instance by avoiding to rely on complex cryptographic primitives.
This paper presented an extensive performance evaluation on the impact of two of such memory protection techniques, Intel software guard extensions (SGX) and AMD secure encrypted virtualization (SEV).
We implemented and deployed a simple, yet representative content-based publish/subscribe system under different hardware configurations.
Our results suggest that AMD SEV is a promising technology: many of our memory-intensive benchmarks run at near native speed.

Additional energy costs can be avoided as long as the system complies with the imposed
restrictions of the hardware-assisted memory protection mechanisms, in particular for Intel SGX.

We hope that our study provides guidance to future system developers willing to implement and deploy privacy-preserving systems exploiting the most recent hardware features.
In order support experimental reproducibility, our code and datasets will be openly released.

\vspace{-5pt}

\section*{Acknowledgments}
The authors would like to thank Christof Fetzer for the discussions on hardware-assisted memory protection mechanisms.
The research leading to these results has received funding from the European Union's Horizon 2020 research and innovation programme under the LEGaTO Project (\href{https://legato-project.eu/}{legato-project.eu}), grant agreement No~780681.

{
\footnotesize
\bibliographystyle{IEEEtranN}
\bibliography{biblio}

}

\end{document}